\documentclass[twocolumn,showpacs,PRL,amsmath,amssymb]{revtex4}
\usepackage{picins}
\usepackage{dcolumn}
\usepackage{bm}
\usepackage{epsfig}

\begin{document}

\title{Bulk evidence for single-gap $s$-wave superconductivity in the intercalated graphite 
superconductor C$_6$Yb }
\author{Mike Sutherland$^{1}$, Nicolas Doiron-Leyraud$^{2}$,
Louis~Taillefer$^{2,3}$,Thomas Weller$^{4}$, Mark Ellerby$^{4}$, S.S. Saxena$^{1}$}

\affiliation{$^1$Cavendish Laboratory, University of Cambridge, J.J. Thomson Ave, Cambridge CB3 OHE, UK\\
$^2$Regroupement Qu\'{e}b\'{e}cois sur les Mat\'{e}riaux de Pointe, D\'{e}partement de physique, Universit\'{e} de Sherbrooke, Sherbrooke, Qu\'{e}bec, Canada\\
$^3$Canadian Institute for Advanced Research, Toronto, Ontarion, Canada\\
$^4$Department of Physics and Astronomy, University College London, Gower Street, London, UK}


\begin{abstract}
 We report measurements of the in-plane electrical resistivity $\rho$ and the thermal conductivity 
 $\kappa$ of the intercalated graphite superconductor C$_6$Yb to temperatures as low as $T_c$/100. 
 When a field is applied along the $c$-axis, the residual electronic linear term $\kappa_0/T$ 
 evolves in an exponential manner for $H_{c1} < H < H_{c2}$. This 
 activated behaviour 
establishes the order parameter as unambiguously $s$-wave, and rules out the possibility of multi-gap
or unconventional superconductivity in this system.

 \end{abstract}

\pacs{74.70.Wz, 74.25.Fy, 74.25.Op}
\maketitle

Carbon is a remarkably versatile element -- in its pure form it may exist as an electronic insulator,
semiconductor or semimetal depending on its bonding arrangement. When dopant atoms are introduced,
superconductivity may be added to this list, observed in graphite 
\cite{Hannay,Koike}, fullerenes \cite{Hebard} and even diamond \cite{Ekimov}. Superconductivity 
in doped carbon was first discovered in the graphite intercalate compounds (GICs), materials 
composed of sheets of carbon separated by layers of intercalant atoms. The first of these compounds 
were intercalated with alkali atoms, and had 
modest transition temperatures of 0.13-0.5 K \cite{Hannay}. The recent discovery of $T_c's$ two 
orders 
of magnitude higher than this in C$_6$Yb \cite{Weller} and C$_6$Ca \cite{Weller,Emery} has however
refocused attention on this intriguing family of compounds. 

 The effects of the intercalant atoms in the GICs are two-fold: they dramatically change the 
 electronic properties of the host graphite lattice by both increasing the separation of the 
 carbon sheets, as well as contributing charge carriers. This causes the two-dimensional 
 graphite $\pi^*$ bands to dip below the Fermi 
 level. The graphite interlayer band, previously unoccupied, also crosses the new Fermi level, 
 contributing three dimensional, free-electron like states located between the carbon sheets. This 
 new interlayer band hybridises strongly with the $\pi^*$ bands, and its occupation appears to be 
 linked with the occurrence of superconductivity in the GICs \cite{Csanyi}. 

There are still several fundamental questions remaining about superconductivity in the GICs, 
especially in C$_6$Yb and C$_6$Ca, where little experimental data 
exists. The pairing mechanism is unresolved, with speculation 
ranging from a conventional route involving the intercalant phonons \cite{Mazin, Mazin2, Calandra}
to superconductivity via acoustic plasmons \cite{Csanyi}. 

Early theoretical studies motivated by the alkali-metal GICs 
\cite{Al-Jishi,Al-Jishi2} emphasized a two-gap model for the superconducting state. In this picture, 
superconductivity arises from coupling between intercalant $s$ electrons and the graphite 
$\pi$ electrons bands, with gaps of different magnitudes existing on different sheets of the Fermi surface. 
Such a scenario is plausible, as there are notable similarities between the GICs and MgB$_2$ 
\cite{Csanyi,Konsin}, a known multi-gap superconductor. Indeed, some aspects of graphite intercalate
superconductivity can be understood by this two-gap phenomenology, however there is little direct 
evidence to support this picture. 

The experimental starting point for addressing these issues is to establish the 
superconducting order parameter. In C$_6$Yb and C$_6$Ca, this task is
complicated by the difficulties in materials preparation arising from the standard vapour 
transport process used for 
intercalation. This process typically yields samples which have a shell of fully intercalated 
material surrounding a core of unintercalated graphite. In addition, both compounds are extremely 
sensitive to air, and their surfaces rapidly deteriorate if left exposed. A recent study of 
penetration depth \cite{Lamura} on C$_6$Ca suggested that the superconductivity was $s$-wave, however 
this technique is extremely surface sensitive, and the results 
were dependent on surface treatment. With these considerations in mind, we turn to 
measurements of bulk thermodynamic properties to probe the superconducting state. 

The technique of thermal conductivity is ideally suited to the study of these materials. It is 
sensitive only to delocalized states, and in highly conductive systems such as C$_6$Yb the majority of
the heat transport at low temperatures is provided by electrons, allowing us to easily separate out 
electronic and phononic contributions to the heat current. Most importantly, thermal conductivity is a 
$\emph{bulk}$ probe, only marginally affected by small concentrations of impurity phases. 

In this Letter we report measurements of low temperature thermal conductivity ($\kappa$) in C$_6$Yb, 
which we use to establish the nature of the superconducting order parameter. The behaviour of 
$\kappa$ as the 
superconducting state is suppressed with a magnetic field shows an activated dependence, clear 
evidence of $s$-wave superconductivity with a single gap energy scale. 

Thermal transport was measured down to 60 mK in a dilution refrigerator using a one heater, two 
thermometer 
steady state technique. Magnetic fields from 0 to 1 T were applied 
parallel to the $c$-axis and perpendicular to the in-plane heat current.  For measurements of 
$\kappa(T)$ at constant field, the sample was cooled in field from T $>$ T$_c$ to maximize
homogeneity of the vortex lattice. 

Our samples were prepared by intercalating very pure highly oriented pyrolytic graphite (HOPG) 
using 
the vapour transport process described elsewhere \cite{Weller}. In 
order to obtain samples consisting mainly of intercalated material, we first cleaved the graphite 
along the $ab$-plane, and then took thin bars from the sides. The resulting samples were rectangular 
platelets of dimensions approximately 1 mm $\times$ 0.5 mm $\times$ 100 $\mu$m. Good quality electrical 
contacts were made using Dupont silver paint applied directly on the surface after cleaving, with 
all handling and mounting done in a glove box under flowing He in order to preserve the quality of
the samples.

       \begin{figure} \centering
              \resizebox{8cm}{!}{
              \includegraphics[angle=0]{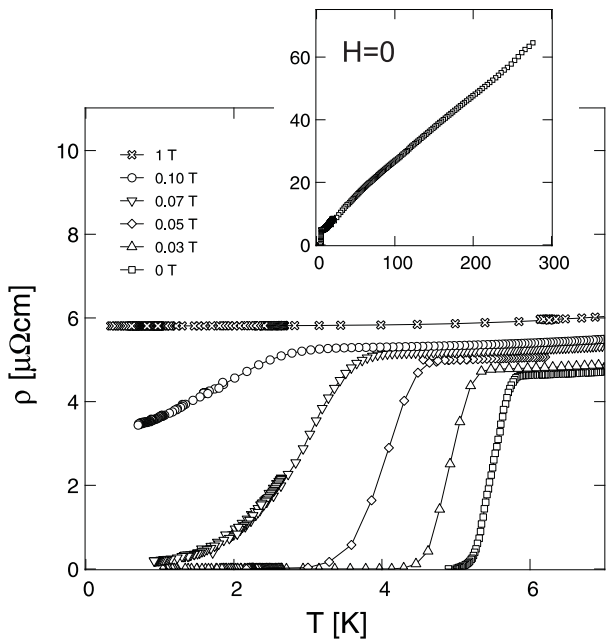}}
              \caption{\label{fig:res} In plane resistivity for C$_6$Yb with
              $H$ $\parallel$ $c$. For $H$ = 0 T the superconducting transition is sharp with 
              T$_c$ = 5.4 $\pm$
              0.4 K but broadens with applied field. For $H$ = 1 T superconductivity is entirely 
              suppressed, revealing a metallic normal state with a residual resistivity
              $\rho_0$ = 5.8 $\mu\Omega$cm.}
              
      \end{figure}
    
Figure 1 shows the in-plane resistivity of C$_6$Yb in both zero field (inset) and 
as a function of field applied along the $c$-axis (main panel). The residual 
resistivity $\rho_0$, is observed to be 4.5 $\mu\Omega$cm by 
extrapolating the zero 
field curve, and slightly larger than this using the 1 T curve. The magnetoresistance is 
comparatively weak in
C$_6$Yb \cite{note2}, only 30 \% by 1 T at 4~K, compared to a factor of 100 increase
by only 0.2 T in pure graphite at $T$ = 5~K \cite{Du}. From the magnitude of 
$\rho_0$ we may 
estimate the electronic mean 
free path $\ell_e$, assuming k$_F$ $\sim$ 0.5 \AA$^{-1}$ \cite{Al-Jishi} we get $\ell_e$ $\simeq$ 
1000 \AA \hspace{4pt}at low temperatures in zero field.

       \begin{figure} \centering
              \resizebox{8cm}{!}{
              \includegraphics[angle=0]{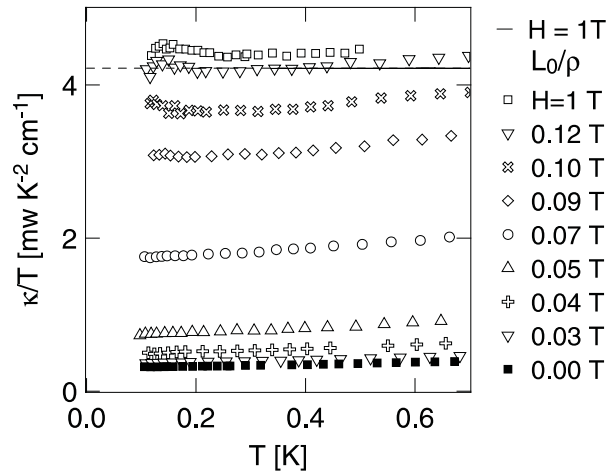}} 
              \caption{\label{fig:2} Low temperature $\kappa$ of C$_6$Yb as
              a function of applied field, with 
              \textbf{J}$\parallel$
              $ab$ and $H$ $\parallel$ $c$. The black line is
              the normal state Wiedemann-Franz law expectation estimated from the resistivity in H = 1 T.}
      \end{figure}

 Figure 2 shows the dependence of the thermal conductivity on field for 
 $H$ $\parallel$ $c$-axis. By plotting $\kappa/T$ vs. $T$ we see that $\kappa/T$ is almost constant, 
 reflecting a dominant electronic contribution. Phonons are responsible for the small slope, which 
 has negligible
impact of the field dependence of the residual linear term, the subject of this paper.
The rise in $\kappa$ with $H$ is gradual at first, then rapidly approaches the normal 
 state value as $H_{c2}$ is reached. By 1~T the sample is fully in the normal state, 
 and a comparison to the electrical resistivity via the Wiedemann-Franz law is shown by 
 the dashed line. In the normal state we see that this law is obeyed to within 5 $\%$, as expected 
 for a metal in the elastic scattering regime.

 Extrapolating the zero field data to $T$ = 0 in Fig.2 yields a small residual linear term 
 $\kappa_0/T$ $\simeq$ 0.3~mWK$^{-2}$cm$^{-1}$. It is tempting to attribute this as arising from nodal quasiparticles 
 in the superconducting state \cite{Durst}, as observed for example in $d$-wave superconductors 
 such as the high-T$_c$ cuprates, where $\kappa_0/T$ = 1.41~mWK$^{-2}$cm$^{-1}$ for 
 overdoped Tl-2201 with T$_c$ = 15 K \cite{Proust}. An estimate of the 
 magnitude 
 of the linear
 term however makes this scenario unlikely. In a nodal superconductor the size of $\kappa_0/T$ 
 is determined by ratio of the quasiparticle velocities parallel ($v_2$) and perpendicular ($v_F$)
 to the Fermi surface near the nodes \cite{Durst}. In a two dimensional $d$-wave superconductor with a gap
 maximum $\Delta_0$ and a density
 of $n$ planes per unit cell of height $c$ we may write

 \begin{equation}
 \frac{\kappa_0}{T} = \frac{k_B^2}{6} \frac{n}{d}k_F\left(\frac{v_F}{\Delta_0}\right)
 \label{eq:Durst}
 \end{equation}

 \noindent assuming $v_F$ $\gg$ $v_2$ \cite{Hawthorn}. 
 Using an average $v_F$ $\simeq$ 3.4 $\times$ 10$^{7}$ cm/s \cite{Al-Jishi2}, 
 k$_F$ $\sim$ 0.5 \AA$^{-1}$ \cite{Al-Jishi} and $\Delta_0$ = 2.14k$_B$T$_c$ = 1.2 meV 
 we expect $\kappa_0/T$ = 6.1~mWK$^{-2}$cm$^{-1}$ for this system. This is over an order of 
 magnitude larger than 
 what we measure, and on this basis we rule out an unconventional order parameter in 
 C$_6$Yb. We instead interpret the finite $\kappa_0/T$ as arising from 
 inclusions of pure graphite where full intercalation was not successful. The extrinsic 
 origin of the linear term is supported by the fact that in a second sample with a similar 
 residual resistivity ($\rho_0$ = 4.4 $\mu\Omega$cm), $\kappa_0/T$ was seen to be half as large. 
 
 Since thermal conductivity probes the bulk of a sample, we may use $\kappa$ as  
 a robust check of the volume fraction of superconducting material. Since an $s$-wave superconductor 
 has a fully gapped Fermi surface in the superconducting state, we would expect no contribution to 
 $\kappa_0/T$ from the intercalated material. The non-intercalated graphite has a
 residual resistivity of $\sim$ 6 $\mu\Omega$cm \cite{Morelli}, and so conducts heat and charge with
 the same ability as bulk C$_6$Yb in the normal state at low temperatures. Thus, the ratio of
 $\kappa_0/T$ in the superconducting state to that in the field induced normal state should, 
 to first approximation, yield the volume fraction of non-superconducting graphite inclusions.
 
 Performing this simple analysis yields ($\kappa_{0,H=0}/T$)/($\kappa_{0,H=1T}/T$) 
 $\simeq$ 7 $\%$, a relatively small fraction of the sample.
 These arguments demonstrate that good quality, essentially bulk 
 samples of C$_6$Yb can be obtained by carefully selecting material from the edges of an intercalated 
 platelet prepared using the vapour transport process.

       \begin{figure} \centering
              \resizebox{8cm}{!}{
              \includegraphics[angle=0]{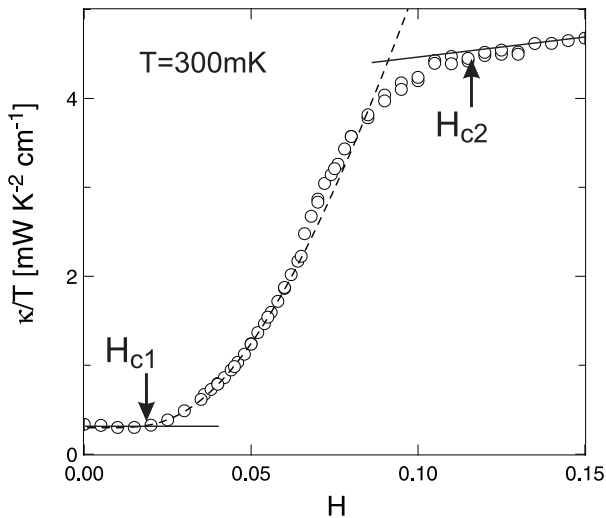}}
              \caption{\label{fig:3} Thermal conductivity of C$_6$Yb at T = 300mK as a function of 
              applied
              field. $H_{c1}$ and $H_{c2}$ are determined as the points where the conductivity
              deviates from a nearly flat $H$ dependence at low and high fields respectively.
              The dotted line is a fit to the expected behaviour for an $s$-wave
              superconductor between $H_{c1} < H < H_{c2}$/2 \cite{Vinen}.}
              
      \end{figure}

We now turn to analysing the field dependence of $\kappa$. Fig. 3 
shows $\kappa/T(H)$ with $T$ held constant at 300 mK. Given the small phonon 
contribution 
evident from the slope of the data inFig. 2, the conductivity in the present figure is largely
due to electrons. Starting from $H$ = 0 we see that $\kappa/T$ is close to zero, and essentially flat 
up to $H$ 
$\simeq$ 0.025 T with a small conductivity arising from the
non-superconducting graphite regions. For higher fields 
a sudden increase in $\kappa/T$ is observed, which we interpret as the onset of the vortex regime 
at $H > H_{c1}$. This agrees reasonably well with estimates of $H_{c1}$ = 0.04 T at low temperatures
from magnetization measurements \cite{Weller}.

As the field is further increased, the conductivity evolves in an exponential manner,
precisely what is expected for transport in the mixed state of an $s$-wave superconductor. As 
vortices first enter the sample at $H > H_{c1}$ the only quasiparticle states at $T \ll T_c$ 
are those associated within 
the vortex cores \cite{Caroli}. When the vortices are far apart, these states remain localized, and 
are unable to contribute to heat transport. Increasing the field decreases the intervortex spacing 
$d \sim \sqrt{\Phi_0/B}$ and the states begin to overlap, forming dispersive bands which yield a 
conductivity 
that grows exponentially with the ratio of the vortex spacing to the coherence length $d$/$\xi$,
$\kappa$ $\propto$ $\sqrt{H}$exp(-$\alpha\sqrt{H_{c2}/H}$) where $\alpha$ is a constant. This dependence is 
readily observed in simple $s$-wave superconductors such as Nb \cite{Vinen} for $H_{c1} < 
H < H_{c2}$/2, 
and is much different than that in nodal superconductors, where the conductivity is observed to 
increase as 
$\sqrt{H/H_{c2}}$ \cite{Proust}. A fit of our data to the
simple $s$-wave form is shown Fig. 3, and the good agreement forms the 
central result of our work: the 
evolution of the electronic conductivity of C$_6$Yb is approximately exponential with applied field, 
providing the first verification of $s$-wave superconductivity in intercalated graphite using a bulk 
thermodynamic technique. 

At still higher fields, the conductivity rolls off and eventually 
saturates as the sample enters the normal state. From the data in Fig. 3 we estimate this
to occur at $H \simeq$ 0.12 T, in good agreement with estimates of $H_{c2}$ =
0.11 T from magnetization measurements \cite{Weller}. With $H_{c2}(\parallel$$c$)
= 0.12 T we estimate $\xi_{ab}$ $\simeq$ 525 \AA $\sim$ $\ell_{e}$, 
which places C$_6$Yb in the dirty limit. This observation is consistent with the fact that the rise 
in conductivity with field is not as dramatic as in clean Nb \cite{Lowell}, but closely 
resembles that observed in dirty limit Nb \cite{Wasim} and metal alloy superconductors 
\cite{Willis,Parks}.

       \begin{figure} \centering
              \resizebox{8cm}{!}{
              \includegraphics[angle=0]{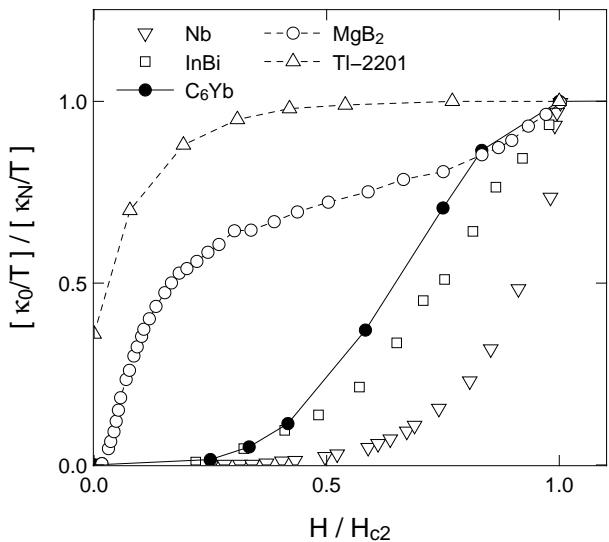}}
              \caption{\label{fig:4}The normalized residual linear term $\kappa_0/T$ of sample B 
              plotted as a function of H/H$_{c2}$, with the small contribution from graphite
              impurities subtracted off. 
              For comparison, similar low-temperature data for the clean $s$-wave superconductor 
              Nb \cite{Lowell}, the dirty $s$-wave superconducting alloy InBi
              \cite{Willis}, the multi-gap superconductor MgB$_2$ \cite{Sologubenko} 
              and an overdoped sample of the $d$-wave superconductor Tl-2201 \cite{Proust} are 
              plotted alongside.}
              
      \end{figure}
 
 In addition to confirming $s$-wave superconductivity, the activated behaviour observed in Fig. 3 
 rules out the scenario of multi-gap superconductivity originally proposed for the GIC's 
 \cite{Al-Jishi}. In Fig. 4 we compare C$_6$Yb to other type-II superconductors by plotting the 
 normalized value of $\kappa_0/T$ 
 versus applied field. The similarity between C$_6$Yb and the alloyed 
 superconductor InBi with T$_c$ = 4.0 K and $H_{c2}$ = 0.07 T \cite{Willis} is 
 striking. Both curves are exponential with field at low H, crossing over to a roughly linear 
 behaviour 
 closer to $H_{c2}$ as expected for $s$-wave superconductors in the dirty limit \cite{Caroli2}. 
 The clean limit
 case observed in pure Nb \cite{Lowell} is shown for contrast. These three curves 
 are very different from the behaviour of the archetypal multi-band superconductor 
 MgB$_2$ \cite{Sologubenko}, or the archetypal $d$-wave superconductor Tl-2201 \cite{Proust}.
  
 In the multiband scenario, gaps of different magnitudes are associated with the $\pi$ and $\sigma$ 
 bands. Such a situation is expected for instance when one band has strong pairing and induces 
 superconductivity in the other by Cooper pair tunnelling \cite{Nakai}, or when electron-phonon
 coupling is significantly different for different bands. Applying a field rapidly 
 delocalizes quasiparticles states confined within the vortices associated with the smaller gap band, 
 while those states associated with the larger gap band delocalize more slowly. This gives rise to the 
 rapid increase in conductivity at low fields and relatively flat dependence at higher fields 
 \cite{Kusunose} evident for in the MgB$_2$ \cite{Sologubenko} data shown in Fig. 4. Our own data 
 thus allows us to rule out any sizable difference between the size 
 of the gaps associated with each band in C$_6$Yb, and suggests a single gap energy scale for the 
 electrons as in the conventional $s$-wave scenario.
 
 In summary we have used bulk measurements of the thermal conductivity $\kappa$ to definitively
 establish $s$-wave superconductivity in C$_6$Yb, and rule out an order parameter with nodes. 
 The activated behaviour of $\kappa_0/T$ also 
 strongly suggests that the pairing state is isotropic, very similar to elementary type-II 
 superconductors in the dirty limit.
 It will interesting to confirm these results on other members of the intercalate family with 
 complementary techniques, although it seems likely that other GICs will share 
 similar superconducting properties.

We would like to thank Robert Smith, Sibel \"{O}zcan and Gil Lonzarich for useful discussions. 
M.S. and N.D-L acknowledge 
support from an NSERC of Canada postdoctoral fellowship, and L.T. acknowledges support from
a Canada Research Chair. This research was funded by EPSRC, NSERC and the CIAR.


\begin{thebibliography}{30}
\bibitem[{not()}]{note2}
\bibinfo{note}{It is probable that all or most of this magnetoresistance arises
  from small inclusions of pure graphite, estimate to make up a few percent of
  our sample in the following paragraphs.}
\end{thebibliography}

\end{document}